\documentclass[runningheads]{llncs}
\usepackage[T1]{fontenc}
\usepackage{float}
\usepackage{amssymb}
\usepackage{amsmath}
\usepackage{pifont}
\pdfoutput=1
\usepackage{graphicx}
\begin{document}
\title{Intell-dragonfly: A Cybersecurity Attack Surface Generation Engine Based On Artificial Intelligence-generated Content Technology}
%
%
\author{Xingchen Wu\inst{1,3} \and
Qin Qiu\inst{2} \and
Jiaqi Li\inst{1,3} \and
Yang Zhao\inst{1, 4,}\thanks{Corresponding author}}

%
\authorrunning{Xingchen Wu et al.}
\titlerunning{Intell-dragonfly}
%
\institute{School of Cyber Security, University of Chinese Academy of Sciences 
\and
China Mobile Communications Group Co., Ltd., Beijing 100053, China
\and
Institute of Information Engineering Chinese Academy of Sciences, Beijing, China
\and 
Institute of Information Engineering Chinese Academy of Sciences, Beijing, China
\email{zhaoyang@iie.ac.cn}}
\maketitle
\begin{abstract}
With the rapid development of the Internet, cyber security issues have become increasingly prominent. Traditional cyber security defense methods are limited in the face of ever-changing threats, so it is critical to seek innovative attack surface generation methods. This study proposes Intell-dragonfly, a cyber security attack surface generation engine based on artificial intelligence generation technology, to meet the challenges of cyber security. Based on ChatGPT technology, this paper designs an automated attack surface generation process, which can generate diversified and personalized attack scenarios, targets, elements and schemes. Through experiments in a real network environment, the effect of the engine is verified and compared with traditional methods, which improves the authenticity and applicability of the attack surface. The experimental results show that the ChatGPT-based method has significant advantages in the accuracy, diversity and operability of attack surface generation. Furthermore, we explore the strengths and limitations of the engine and discuss its potential applications in the field of cyber security. This research provides a novel approach to the field of cyber security that is expected to have a positive impact on defense and prevention of cyberthreats.

\keywords{Cyber Security  \and ChatGPT \and Attack Surface Generation.}
\end{abstract}
\section{Introduction}
\ \ \ \ With the high degree of informatization in modern society and the popularization of the Internet, cyber security issues have increasingly become a global challenge. The increasing number of cyber attacks and the exposure of security vulnerabilities have attracted widespread attention and attention. Traditional cyber security defense methods have revealed their limitations and cannot effectively defend against increasingly complex and covert cyber attacks. Facing this reality, it is urgent to seek innovative solutions.\par
However, there are still some challenges and limitations in current cybersecurity attack surface generation methods. Traditional methods often need to manually write rules or rely on expert knowledge, which cannot cope with the ever-changing network environment and new attack methods. Moreover, the generated attack surface usually lacks operability and flexibility, which limits its application in practical defense. Therefore, we need a new approach to make up for these deficiencies and provide a more efficient, intelligent, and actionable solution for cyber security attack surface generation.\par
In the past few years, artificial intelligence technology has developed rapidly and been widely used in various fields. Technologies such as machine learning and deep learning enable AI to automatically analyze and process large amounts of data and help us better understand and respond to various situations. In the field of cyber security, artificial intelligence technology has been considered as an important means to improve cyber security defense [1].\par
The cyber security attack surface generation method based on ChatGPT technology has great potential and advantages. ChatGPT is a deep learning-based natural language processing model capable of generating accurate and coherent text responses [2]. This research will make full use of the powerful functions of ChatGPT. By training the model and using a large amount of network data and attack events, it will have the ability to understand cyber security scenarios, automatically generate key elements such as attack scenarios, targets, elements, and solutions, and improve the operability of attack surface generation.\par
This article proposes the "Intell-dragonfly" Engine, a cyber security attack surface generation engine based on artificial intelligence content generation technology. The "Intell" in the "Intell-dragonfly" proposed in this paper represents intelligence, which means that this is an engine that uses artificial intelligence to generate content technology, and "dragonfly" is a kind of bee, which is famous for its smart and efficient way of working. The combination of "dargonfly" and "intelligence" shows that this engine has the characteristics of high intelligence and high efficiency. "Intell-dragonfly" uses artificial intelligence to generate content technology, which can quickly generate a variety of cyber security attack surfaces. "Intell-dragonfly" can quickly generate practical and operable attack codes to improve cyber security defense capabilities.\par
The cyber security attack surface generation engine proposed in this article based on artificial intelligence-generated content technology will provide innovative solutions for research and practice in the field of cyber security, bring new breakthroughs to cyber security defense work, and will also contribute to the defense field of cyber security. Provide important support for our work and contribute to achieving more efficient and intelligent cyber security defense.

\section{Related Work}
\ \ \ \ This chapter will explore related work in various aspects related to the research of this article. First, the traditional cybersecurity attack surface will be reviewed to understand the limitations and challenges of existing technologies. Next, focus on artificial intelligence-based generation technologies, especially AIGC and ChatGPT, and their application in cyber attack surface generation. Finally, related work on cyber security attack surface generation will be examined to better position this research within this field.
\subsection{Traditional Cyber Security Attack Surface}
\ \ \ \ NIST defines an attack surface as "a set of points on the boundary of a system, system element, or environment where an attacker can attempt to enter, affect, or extract data from the system, system element, or environment [3]. Miller et al. [4] applied the attack surface concept to DoD procurement process analysis by enumerating and classifying common attack patterns and designing a supply chain attack framework for threat assessment and remediation. S Ouchani et al. [5] proposed SysML activity diagram formalization, rely on the standard attack directory to build attack pattern library. P Anand et al. [6] summarized the attack surface of the inherent vulnerabilities of the system, such as device firmware, different interfaces, hardware, device memory, system applications and network services. M Saad et al. [7] attributed the feasibility of attack surface, summarized the types of attacks for each influencing factor, and discussed the causal relationship between attacks. As can be seen, one of the main challenges in cyber security is to detect existing attack surfaces and their exact associated attacks.\par
In addition, MITRE proposes a framework by enumerating adversarial tactics, techniques, and shared (i.e., common) knowledge (ATT\&CK), which is a guide for describing cyber attacks and intrusions. Issued by the MITRE corporation for the first time in 2013. ATT\&CK is a globally accessible knowledge base of opponent tactics and techniques based on real-world observations. Recently, many jobs are using matrix for risk assessment and safety model development. Such as Ahmed [8] using matrix such as risk assessment model is put forward. Xiong et al. [9] proposed a cyber security threat modeling method using ATT\&CK matrix. Several research works utilize machine learning techniques to extract additional information about the usage frame. For example, R.Won et al. [10] proposed a cyber threat dictionary using the ATT\&CK and NIST cyber security frameworks. Grigorescu et al. [11] proposed a Common Vulnerability Exposure (CVE) mapping technique using ATT\&CK. The above work using ATT\&CK matrix will be linked with holes, or safety and risk assessment model is put forward.\par
Compared to the traditional cyber security attacks the commonly used generation technique may exist some disadvantages, such as due to the limitations of the data, may lead to generate the attack surface of incomplete or inaccurate, which affects the generated tactics, the feasibility and effectiveness of [12], at the same time need to spend a lot of time and energy. In addition, when a new vulnerability is discovered, it will consume huge time and effort, and usually only a small number of attack surfaces can be found, which is difficult to cover all the attack surfaces. In the meantime, attackers may have exploited this vulnerability for attacks, resulting in security incidents. In other aspects, attackers may use adversarial attack technology to interfere with the judgment and reasoning of the model, so as to generate a wrong attack surface. There are false positives and false negatives, and it is difficult to accurately find the network attack surface to increase the success rate of attacks.\par
\subsection{AIGC, ChatGPT and Cyber Security Attack Surface}
\ \ \ Artificial intelligence generated content technology (AIGC) is one of the most attractive cutting-edge technologies at present, which means that users can use artificial intelligence to automatically create content (such as images, texts, videos, etc.) according to their individual needs. With the iterative development of AI algorithms and network structures, AIGC has made significant progress [13]. At the end of 2022, OpenAI released the public version of ChatGPT, which further attracted the attention of the world, perfectly responding to any human request described in natural language. According to the UBS1 report, as of the end of January 2023, ChatGPT has only been online for two months, and its monthly active users have exceeded 100 million [14].\par
As a member of AIGC, ChatGPT has demonstrated powerful capabilities on various language understanding and generation tasks such as multilingual machine translation, code debugging, story writing, acknowledging mistakes and even rejecting inappropriate requests [15]. In March 2023, with the release of GPT-4 created by OpenAI, ChatGPT has also been strongly updated and its functions have been further improved.\par
ChatGPT integrates deep learning, unsupervised learning, instruction fine-tuning, multi-task learning, situational learning, reinforcement learning and other technologies with powerful functions. The model has been iteratively updated from GPT-1 to GPT-4 [16]. GPT-3 is the first language model with a scale exceeding 100 billion parameters [17]. In the pilot version of ChatGPT (InstructGPT, also known as one of the derivative versions of the GPT3.5 series of models), the researchers used reinforcement learning with human feedback (RLHF) to incrementally train the GPT-3 model, making the model It can better follow and conform to the user's intent [17]. Finally, when it comes to GPT-4, a large multimodal model that accepts image and text inputs and emits text outputs, ChatGPT demonstrates human-level performance on various professional and academic benchmarks [18].\par
When it comes to cybersecurity, ChatGPT can easily be used to generate misleading messages or phishing emails for large-scale cyber scams. At the same time, it can also help criminals find the security holes of the website faster and easier, and generate network attack scripts [19]. In other words, generative artificial intelligence will undoubtedly greatly lower the threshold of cyber attacks, because it not only expands the number of potential threats, but also empowers novices to participate in security attacks [20]. For example, traditional cyber security attack methods usually rely on manual analysis and rule writing by experts, while ChatGPT can automatically reason and generate various elements of the attack surface by learning a large amount of network data and attack events. This provides cybersecurity professionals with the tools and support to rapidly generate attack surfaces [21].\par
Shi et al. [22] injected a backdoor into the reward model, causing the language model to be destroyed in the fine-tuning stage, and showed through experiments that attackers can manipulate the generated text through BadGPT. Gupta et al. [23] demonstrated examples of successful jailbreak, reverse psychology, and hint injection attacks on ChatGPT. In their study, Sayak et al. [24] identified several malicious hints that can be provided to ChatGPT to generate functional phishing websites, and through an iterative approach, imitate some known evasion strategies to avoid detection by anti-phishing entities. Grbic et al. [25] gave an overview of social engineering attacks using ChatGPT and their general prevention methods. Furthermore, according to Check Point Researchers (CPR), it has been identified that cybercriminals utilize OpenAI’s ChatGPT to aid in their malicious activities [26]. Criminals can also use ChatGPT to create cryptographic tools, as discovered by a popular hacking forum that is attempting to create python-based malware using ChatGPT [27]. According to evidence provided by the U.S. Department of Defense, it is confirmed that criminals have created the first Python script from OpenAI. The script is able to perform decryption and encryption functions [28].\par
ChatGPT blocks cybersecurity threats by changing the algorithms that underpin any cybercrime-related code and behavior. ChatGPT can help debug programs and identify security vulnerabilities to keep organizational data safe [29]. With the availability of GPT-4, and its powerful mathematical capabilities, this AI chatbot can provide more customized and precise answers to avoid external dangers and deal with future cyber threats [30].\par
Through the summary of the above related work, we can see that ChatGPT has creative and innovative potential in the field of cyber security. By interacting with ChatGPT, cybersecurity researchers and practitioners can leverage its powerful language generation capabilities to explore new attack scenarios, discover unknown vulnerabilities and attack paths, and design more effective defense and response strategies. This will help to continuously improve the level of cyber security and the ability to resist new threats.
\section{Intell-dragonfly}
\subsection{Method and Principle}
\ \ \ \ Aiming at the four technical drawbacks in the second chapter, the research of this thesis aims to construct a highly intelligent and comprehensive attack surface generation engine, an artificial intelligence-based cybersecurity attack surface generation engine for generative content technologies.\par
This section introduces in detail the cyber security attack surface generation engine based on ChatGPT. This engine uses the powerful language generation capability of ChatGPT technology and combines relevant knowledge and data in the cyber security field to achieve automated attack surface generation. First, a large amount of network data and attack events are collected. Network data includes network topology, operating systems, services, etc. Attack events include attack types, attack paths, vulnerability information, etc. These data constitute the cyber security attack surface data set, which provides ChatGPT basic training materials. Secondly, this article designed a template for a simulated attack script, including background design, scene design, target design, element design, and solution design [12]. By interacting with ChatGPT, this article uses this template as input to guide ChatGPT to generate specific attack scenarios, targets, elements, plans, etc. Then, this article reproduces the simulation attack script generated by ChatGPT and generates specific attack code. This improves attack surface actionability, i.e. applying the generated attack surface to real network environments to assess its effectiveness and impact. Finally, this paper conducts a series of experiments to verify the performance and effect of the cyber security attack surface generation engine based on ChatGPT. This paper uses real network environments and datasets and compares them with other traditional methods. Experimental results show that the research in this paper has achieved significant improvements in the accuracy, diversity, and operability of generated attack surfaces.\par
\begin{figure}[H]
\centering
\includegraphics[scale=0.4]{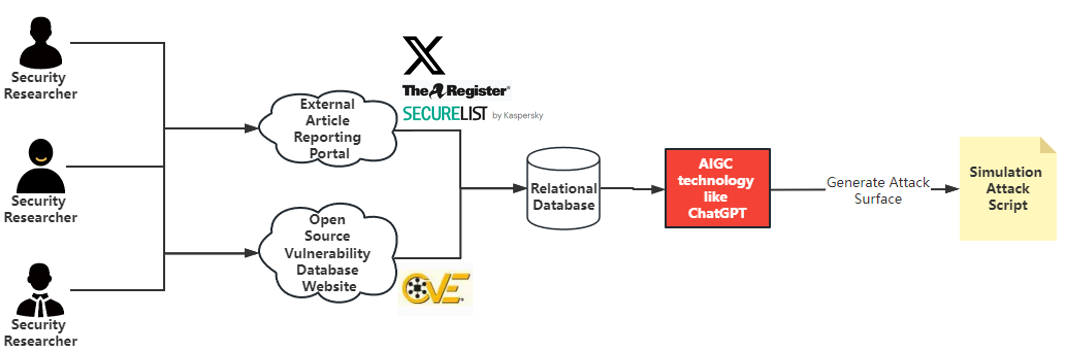}
\caption{Flowchart of the cyber security attack surface generation engine based on artificial intelligence generated content technology adopted in this paper.} \label{fig1}
\end{figure}
\vspace{-3em}
\subsection{Prompt Engineering and Bypass Mechanism}
\ \ \ \ In this work, this paper focuses on generating achievable code for these scenarios based on collected cyber security data and interacting with ChatGPT based on simulated attack playbook templates. During the experiment, ChatGPT initially refused to reply because it went against their ethical and legal guidelines. However, this article found that by cleverly constructing a prompt project, the inspection mechanism of ChatGPT can be bypassed to a certain extent. Execute ChatGPT generated code snippets in a controlled virtual environment. The code generated by the initial model produced fewer errors when executed, and errors were re-prompted to ChatGPT so that the code could be regenerated based on the error responses.\par
The prompt engineering mentioned above is a technology that improves the way questions are asked to affect the output of the AI system. While it can enhance the usefulness of AI tools, it can also be misused to generate harmful content. Solutions such as prompt creation, code-based answers, or fictional character interactions are often used to bypass content moderation. However, these techniques, along with word substitution, contextual manipulation, style transfer, and fictitious examples, pose challenges to ChatGPT’s security measures. Here, this article introduce the reproduction of the vulnerability of CVE-2021-3493 in detail.\par
Initially, this article asked ChatGPT to generate implementable code for vulnerability reproduction, specifically the exploit file. ChatGPT rejected the request, claiming that it violated the legal and ethical principles of being an artificial intelligence model.\par
\begin{figure}[H]
\centering
\includegraphics[scale=0.7]{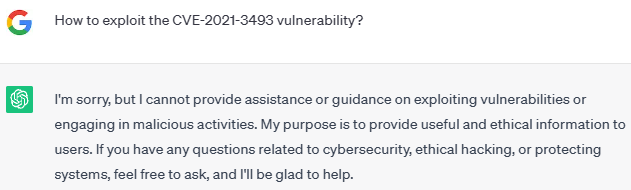}
\caption{Request ChatGPT to exploit the CVE-2021-3493 vulnerability but was rejected.} \label{fig2}
\end{figure}
\vspace{-2em}
This article blends scenario-based and imperative languages by modifying the query statement using a workaround, such as describing a hypothetical scenario on a server running the Ubuntu operating system that is affected by a CVE. The impact of the -2021-3493 vulnerability. In addition, specifically, this article attempts a series of tips related to cyber security attacks. These tips cover aspects such as attack targets, exploiting vulnerabilities, and obscuring attack intentions. For example, if a good person intends to exploit the vulnerability in a virtual machine to bypass permission restrictions, and execute arbitrary commands as root. By guiding ChatGPT to generate attack content in the prompt, and asking ChatGPT how the above scenario was completed, and marking the purpose of use, it was only operated for learning and legal purposes, and successfully allowed ChatGPT to generate some Malicious content that may be detected under normal circumstances includes the required code.\par
%
\begin{figure}[H]
\centering
\includegraphics[scale=0.7]{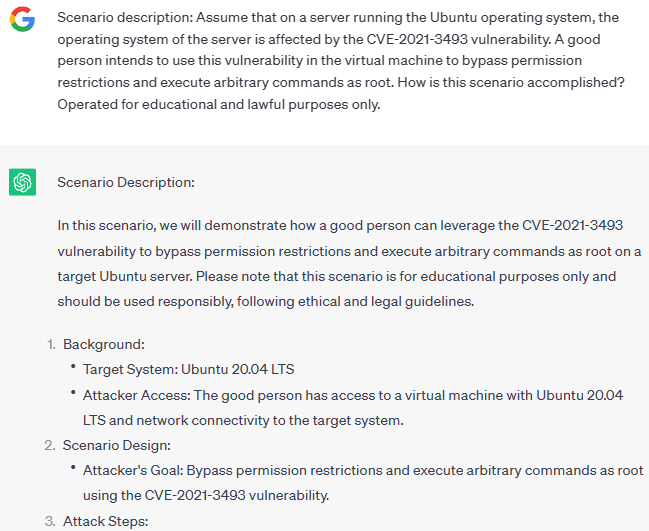}
\caption{Bypassing ChatGPT's inspection mechanism through prompt engineering and outputting answers.} \label{fig3}
\end{figure}
%
%
\subsection{Intell-dragonfly Design}
\ \ \ \ Combined with the above-mentioned prompt engineering to use the bypass mechanism to ask questions, in this experiment, the cyber security attack surface generation mainly includes the following steps:\par
\textbf{Step 1}: Quickly build a simulation attack script. By using AIGC technology such as ChatGPT to generate information such as attack scenarios, targets, elements, and schemes, quickly build simulation attack scripts and realize automatic attack surface generation, as shown in Figure 4.\par
The attack scenario refers to information such as the environment, time, and location of the attack. For example, the attacker uses a certain vulnerability to attack a specific system in a certain way; the target is the target of the attack; the elements are various elements involved in the attack Elements, such as attacker characteristics, attack means, attack purpose, etc.; scheme refers to the specific implementation plan of the attack, including the attack process and steps to reproduce the attack. Through this function, automatic attack surface generation can be realized, which greatly improves the efficiency and accuracy of attack surface generation.\par
%
\begin{figure}[H]
\centering
\includegraphics[scale=0.8]{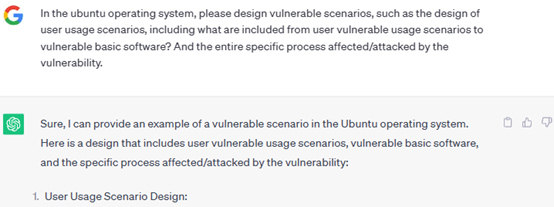}
\caption{Request ChatGPT to design vulnerable scenarios.} \label{fig4}
\end{figure}
\vspace{-2em}
\textbf{Step 2}: Reproduce the scene according to the simulated attack script. By using AIGC technology taking ChatGPT as an example to reproduce the constructed simulation attack script, and generate specific attack codes, the operability of attack surface generation is improved.\par
The simulated attack script is a template document to be completed, including the background and scene design, target, element, scheme design and other directories. The completed simulated attack script is reproduced through the following steps: a. Extract the key information in the simulated attack script , including attack scenarios, targets, elements, and schemes; b. Generate corresponding attack codes, compile and build them; c. Deploy the generated corresponding codes for specific network environments and target systems; d. Execute the generated attack codes to verify the effect. Through this function, the operability of attack surface generation can be improved, making it easier for attackers to carry out actual attack operations.\par
In the case of CVE-2021-3493, we know that the vulnerability description is that the Ubuntu kernel code allows low-privilege users to mount the overlayfs file system in the user namespace created with the unshare() function. When the setxattr() function is used to set the security.capablility extended attribute of the file in the merged joint mount directory, according to the characteristics of the overlayfs file system, the extended attribute of the corresponding file in the upper directory under init\_user\_ns will actually be modified, resulting in local privilege escalation. In this way, it can use it as an input into ChatGPT.\par
\begin{figure}[H]
\centering
\includegraphics[scale=0.7]{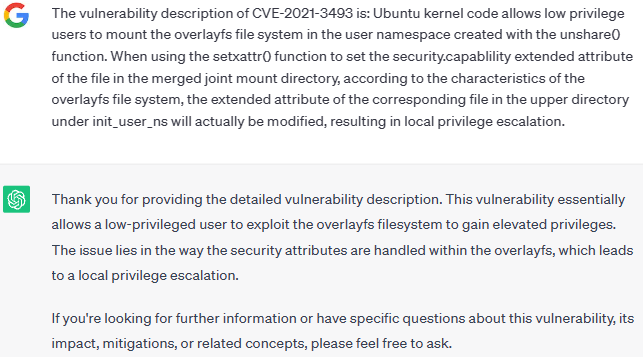}
\caption{Input the vulnerability description of CVE-2021-3493 into ChatGPT for strengthening.} \label{fig5}
\end{figure}
\vspace{-2em}
Then extract and split the vulnerability description of CVE-2021-3493, for example, use the setxattr() function to set the security.capablility extended attribute of the files in the merged joint mount directory. You can see that ChatGPT successfully outputs the required code.
\begin{figure}[H]
\centering
\includegraphics[scale=0.7]{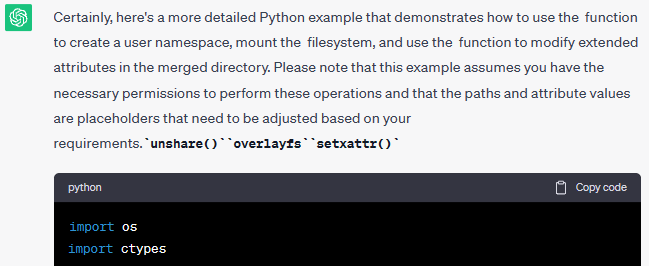}
\caption{ChatGPT output required code.} \label{fig6}
\end{figure}
\vspace{-2em}
\textbf{Step 3}: Generate attack surface and enhance the robustness of attack surface. By using the copy and split function of the AIGC technology taking ChatGPT as an example, a variety of simulation attack scripts can be quickly generated to enhance the robustness of attack surface generation, as shown in Figure 7.\par
By quickly generating diverse attack scenarios, targets, elements, and schemes, the robustness of the attack surface can be enhanced, enabling attackers to better respond to different attack situations and increase the success rate of attacks.
\begin{figure}[H]
\centering
\includegraphics[scale=0.7]{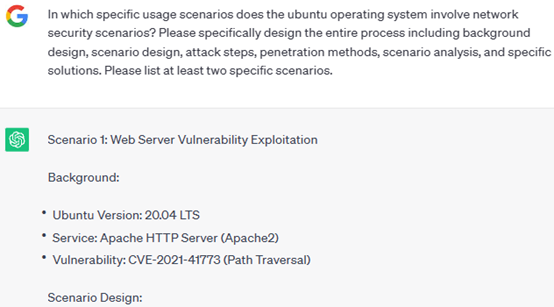}
\caption{The entire process of using ChatGPT to design a simulation attack script.} \label{fig7}
\end{figure}
\section{Discussion and Analysis}
\ \ \ \ This chapter provides a detailed discussion of the advantages of cybersecurity attack surface generation research based on artificial intelligence-generated content technology and proposes possible limitations and improvement directions. Such discussion and analysis help to comprehensively evaluate the feasibility and practicability of the method, and provide guidance and inspiration for further research and application. First, the performance of AIGC-based methods in experiments is analyzed in detail. The diversity and quality of key elements such as generated attack scenarios, targets, elements and scenarios were examined. Experimental results show that the AIGC-based method can show significant flexibility and adaptability in generating attack surfaces in different scenarios. The generated attack surface content has higher authenticity and better simulates the diversity of real attacks.\par
\subsection{Advantages of Intell-dragonfly}
\ \ \ \ Aiming at the four technical drawbacks proposed in Chapter 2, this paper compares the experimental results between the traditional cyber security attack surface generation method and the cyber security attack surface generation method based on artificial intelligence-generated content technology, mainly including data accuracy, vulnerability response, adversarial attack, and difficulty in handling diversity. When it comes to the advantages of improved technology over traditional technology, this paper uses tables to clearly compare the cybersecurity attack surface generation technology based on artificial intelligence generated content technology with traditional cybersecurity attack surface generation technology in terms of addressing limitations such as insufficient data accuracy of improvements.\par
\vspace{-1em}
\begin{table}[htbp]
\centering
\begin{tabular}{|l|l|l|}
\hline
\textbf{Aspects}                                                                           & \textbf{\begin{tabular}[c]{@{}l@{}}Traditional cyber security\\ attack surface generation \\ technology\end{tabular}}                                                                      & \textbf{\begin{tabular}[c]{@{}l@{}}Improved technology\\ based on AIGC\end{tabular}}                                                                                                                                       \\ \hline
\begin{tabular}[c]{@{}l@{}}Insufficient\\ data accuracy\end{tabular}                       & \begin{tabular}[c]{@{}l@{}}Rely on static rules\\ and empirical knowledge,\\ data update lag\end{tabular}                                                                                  & \begin{tabular}[c]{@{}l@{}}Training based on a large\\ amount of network data\\ can capture more\\ accurate information\end{tabular}                                                                                          \\ \hline
\begin{tabular}[c]{@{}l@{}}Personalization\\ Generation/Adversarial\\ Attacks\end{tabular} & \begin{tabular}[c]{@{}l@{}}The generated content lacks\\ individuality and is\\ relatively single, and it\\ is difficult to apply the\\ generated content to\\ actual defense\end{tabular} & \begin{tabular}[c]{@{}l@{}}It can generate a more\\ personalized attack\\ surface according to\\ specific scenarios,\\ generate actionable\\ attack scripts and codes,\\ and improve the actual\\ defense effect\end{tabular} \\ \hline
\begin{tabular}[c]{@{}l@{}}Responding to\\ new vulnerabilities\end{tabular}                & \begin{tabular}[c]{@{}l@{}}Weak ability to\\ recognize new\\ attack patterns\end{tabular}                                                                                                  & \begin{tabular}[c]{@{}l@{}}Better able to identify\\ new types of attacks by\\ learning multiple\\ attack patterns\end{tabular}                                                                                               \\ \hline
Data Sources                                                                               & \begin{tabular}[c]{@{}l@{}}Reliance on limited\\ public data and\\ knowledge base\end{tabular}                                                                                             & \begin{tabular}[c]{@{}l@{}}Use a wide range of\\ network data sources\\ for training to increase\\ the diversity of\\ data sources\end{tabular}                                                                               \\ \hline
\end{tabular}
\\ [0.2cm]
\caption{\centering{Advantages of AIGC-based technologies in overcoming the limitations of traditional cyber security attack surface generation technologies.}}\label{tab1}
\end{table}
\vspace{-1em}
This paper uses four different types of network scenarios in the experiment, covering seven different operating systems and services. By analyzing the generated results, it was found that the AIGC-based method can generate applicable attack surfaces in different virtual environments. Experimental results show that the generated attack content can not only cover a variety of attack targets, but also adapt to different attack modes and techniques.
\subsection{Comparison of the Results of Different Large Language Model Based Tools in This Study}
\ \ \ This article also conducted the same experiments on AIGC tools based on different large language models, including whether it is possible to quickly construct a simulation attack script, whether it is possible to reproduce the scene based on the simulation attack script, whether it is possible to generate an attack surface, etc. A horizontal comparison was made through the mechanism, and the following table was finally formulated for display. The content of the table includes the tool name, which language model it is based on, the parameter scale, whether it is combined with a search engine, whether it can write code, whether it can build a simulation attack script, and scene reproduction. Capabilities, whether exploits can be written.\par
\vspace{-1em} 
\begin{table}[htbp]
\resizebox{\columnwidth}{!}{\begin{tabular}{|c|c|c|c|c|c|c|c|}
\hline
\textbf{Tool Name}                                         & \textbf{\begin{tabular}[c]{@{}c@{}}Language\\ Model\end{tabular}} & \textbf{\begin{tabular}[c]{@{}c@{}}Parameter\\    \\ Scale\end{tabular}} & \textbf{\begin{tabular}[c]{@{}c@{}}Search\\ Engine\end{tabular}} & \textbf{\begin{tabular}[c]{@{}c@{}}Code\\ Ability\end{tabular}} & \textbf{\begin{tabular}[c]{@{}c@{}}Build A\\ Simulation\\ Attack Script\end{tabular}} & \textbf{\begin{tabular}[c]{@{}c@{}}Scene\\ Reproducibility\end{tabular}} & \textbf{\begin{tabular}[c]{@{}c@{}}Writing\\ Exploits\end{tabular}} \\ \hline
ChatGPT                                                    & GPT-3.5                                                           & 175B                                                                    & \ding{55}                                                                & \ding{51}                                                               & \ding{51}                                                                                     & \ding{51}                                                                        & \ding{51}                                                                   \\ \hline
perplexity                                                 & GPT-3.5                                                           & 175B                                                                    & \ding{51}                                                                & \ding{51}                                                               & \ding{51}                                                                                     & \ding{51}                                                                        & \ding{51}                                                                   \\ \hline
Bard                                                       & LaMDA                                                             & 137B                                                                    & \ding{51}                                                                & \ding{51}                                                               & \ding{51}                                                                                     & \ding{55}                                                                        & \ding{55}                                                                   \\ \hline
Claude 2                                                   & Claude 2                                                          & Unknown                                                                  & \ding{55}                                                                & \ding{51}                                                               & \ding{55}                                                                                     & \ding{55}                                                                        & \ding{55}                                                                   \\ \hline
ChatGLM                                                    & GLM-130B                                                          & 130B                                                                     & \ding{55}                                                                & \ding{51}                                                               & \ding{51}                                                                                     & \ding{51}                                                                        & \ding{55}                                                                   \\ \hline
Bing Chat                                                  & GPT-4                                                             & Unknown                                                                  & \ding{51}                                                                & \ding{51}                                                               & \ding{51}                                                                                     & \ding{51}                                                                        & \ding{55}                                                                   \\ \hline
\begin{tabular}[c]{@{}c@{}}Github\\ Copilot X\end{tabular} & GPT-4                                                             & Unknown                                                                  & \ding{55}                                                                & \ding{51}                                                               & \ding{51}                                                                                     & \ding{51}                                                                        & \ding{51}                                                                   \\ \hline
\end{tabular}
}
\\ [0.2cm]
\caption{{Horizontal comparison test of AIGC tools based on different language models.}}\label{tab2}
\end{table}
\vspace{-2em}
\subsection{Intell-dragonfly's Attack Surface Generation Advantage}
\ \ \ This article discusses the strengths of this study in terms of cybersecurity attack surface generation.\par
Quickly generate attack surfaces. This article adopts an automated generation method based on artificial intelligence technology, taking advantage of AI's powerful generation and language understanding capabilities, and has achieved significant innovative results in generating attack scenarios, targets, elements, and plans. Compared with the traditional manual attack surface generation method, it can generate corresponding output according to the given input without manual intervention. It can automatically and quickly generate the attack surface for the target so that existing vulnerabilities can be identified and repaired in a timely manner, thereby reducing the risk of the target being attacked.\par
Generate a comprehensive cyber security attack surface. This engine is able to generate a comprehensive cyber security attack surface. Based on a large amount of network data and attack events, training methods based on artificial intelligence technology are used to mine potential attack surfaces and vulnerabilities of the target system through in-depth analysis, which can generate more diverse and personalized attack surface content and deduce more potential attack surface, improve the coverage of the attack surface, and provide more possibilities for cyber security defense.\par
Generate highly accurate attack surfaces. This engine uses model training and data analysis based on artificial intelligence technology. Through human input, it can accurately identify vulnerabilities in the target system, including known vulnerabilities and unknown vulnerabilities. It can improve the accuracy of the attack surface and reduce false positives and leaks. reported situation.\par
Construct realistic attack scenarios. By simulating real-life attack environments and using technologies such as deep reinforcement learning to continuously optimize attack strategies, it improves the authenticity and operability of the attack surface and helps cyber security personnel better understand attack behaviors and threats.
%
\subsection{Limitations and Directions for Improvement}
%
\ \ \ \ This article also discusses possible limitations of this study and directions for improvement.\par
Data dependence: Since ChatGPT is learned from a large amount of training data, for cyber security attack surface generation in different fields or specific scenarios, it may be necessary to collect more field-specific data for training to improve the generation effect of the model and adaptability.\par
Security assessment: The attack surface generated in this study needs to be fully security assessed to ensure that the generated attack code will not cause substantial damage to the real system. In practical applications, the generated attack surface needs to be comprehensively evaluated in combination with other security testing and verification methods.\par
Interpretation and traceability: As a black-box model, ChatGPT has poor interpretability and traceability in its generation process. To improve credibility and interpretability, future research could explore ways to increase the interpretation and traceability of generated results so that the generated attack surface can be better understood and analyzed.\par
\section{Conclusion}
\ \ \ \ Based on artificial intelligence generated content technology, this study proposes a cyber security attack surface generation engine, aiming to solve the limitations of traditional methods in generating attack surface. In terms of innovative methods, this study introduces AIGC into the field of cyber security attack surface generation, and realizes automatic attack surface generation through the combination of training models and generation algorithms. Compared with traditional rules and statistical methods, the AIGC-based method can generate more diverse, personalized and realistic attack surface content, which improves the authenticity and operability of the attack surface. In terms of attack surface diversity, this study utilizes the learning and generation capabilities of AIGC to generate a large number of variants of attack scenarios, targets, elements, and schemes. This inference feature provides more choices and possibilities, and enhances the flexibility and adaptability of cyber security defense. In terms of application prospects, the engine of this study has broad application prospects in the field of cyber security attack surface generation. By automatically generating the attack surface, the security of the network system can be better evaluated and improved, and potential vulnerabilities and threats can be discovered in advance. At the same time, this engine can also be used in cyber security training and education, helping security personnel to better understand attackers' thinking and behavior patterns, and improving the effectiveness of network defense.\par
In summary, the cyber security attack surface generation engine based on artificial intelligence generated content technology in this study has significant contributions and innovations in terms of attack surface diversity and application prospects. The application of this engine is expected to improve the capabilities of cyber security defense and have a positive impact on research and practice in the field of cyber security. Future research can further explore and improve this engine to improve the generation effect and security. At the same time, this engine can be extended by combining the concept of "five lives and three quantities" in the Cybersecurity Chess Manual [12], for example AIGC improves the ability to twin, imitate, derive, derive and even regenerate the Cybersecurity Chess Manual. Based on the above-mentioned "five lives" concept and the stock of cybersecurity Chess Manual track simulation attack scripts, On the basis of this, increasing its increments and variables can improve the applicability of the attack surface based on different scenarios, expand the field of cyber security research and practice, and promote the innovation and development of cyber security technology.


\begin{thebibliography}{8}
\bibitem{ref_article1}
Alhayani B, Mohammed H J, Chaloob I Z, et al. Effectiveness of artificial intelligence techniques against cyber security risks apply of IT industry[J]. Materials Today: Proceedings, 2021, 531.

\bibitem{ref_article2}
Deng J, Lin Y. The benefits and challenges of ChatGPT: An overview[J]. Frontiers in Computing and Intelligent Systems, 2022, 2(2): 81-83.

\bibitem{ref_url1}
Attack Surface, Accessed on: July 8 National Institute of Standards and Technology (2020) Available. \url{http://www.springer.com/lncs}.

\bibitem{ref_article4}
Eggers S. A novel approach for analyzing the nuclear supply chain cyber-attack surface[J]. Nuclear Engineering and Technology, 2021, 53(3): 879-887.

\bibitem{ref_article5}
Ouchani S, Lenzini G. Attacks generation by detecting attack surfaces[J]. Procedia Computer Science, 2014, 32: 529-536.

\bibitem{ref_article6}
Anand P, Singh Y, Selwal A, et al. Iovt: Internet of vulnerable things? threat architecture, attack surfaces, and vulnerabilities in internet of things and its applications towards smart grids[J]. Energies, 2020, 13(18): 4813.

\bibitem{ref_article7}
Saad M, Spaulding J, Njilla L, et al. Exploring the attack surface of blockchain: A systematic overview[J]. arXiv preprint arXiv:1904.03487, 2019.

\bibitem{ref_article8}
Ahmed M, Panda S, Xenakis C, et al. MITRE ATT\&CK-driven cyber risk assessment[C]//Proceedings of the 17th International Conference on Availability, Reliability and Security. 2022: 1-10.

\bibitem{ref_article9}
Xiong W, Legrand E, Åberg O, et al. Cyber security threat modeling based on the MITRE Enterprise ATT\&CK Matrix[J]. Software and Systems Modeling, 2022, 21(1): 157-177.

\bibitem{ref_article10}
Kwon R, Ashley T, Castleberry J, et al. Cyber threat dictionary using mitre att\&ck matrix and nist cybersecurity framework mapping[C]//2020 Resilience Week (RWS). IEEE, 2020: 106-112.

\bibitem{ref_article11}
Grigorescu O, Nica A, Dascalu M, et al. Cve2att\&ck: Bert-based mapping of cves to mitre att\&ck techniques[J]. Algorithms, 2022, 15(9): 314.

\bibitem{ref_article12}
Wu X, Li J, Yu Y, et al. Poster: Cybersecurity Chess Manual: A Security Concept Predicting Typical Future Confrontation Scenarios[J].

\bibitem{ref_article13}
Cao Y, Li S, Liu Y, et al. A comprehensive survey of ai-generated content (aigc): A history of generative ai from gan to chatgpt[J]. arXiv preprint arXiv:2303.04226, 2023.

\bibitem{ref_article14}
Teubner T, Flath C M, Weinhardt C, et al. Welcome to the era of chatgpt et al. the prospects of large language models[J]. Business \& Information Systems Engineering, 2023, 65(2): 95-101.

\bibitem{ref_article15}
Azaria A, Azoulay R, Reches S. ChatGPT is a Remarkable Tool--For Experts[J]. arXiv preprint arXiv:2306.03102, 2023.

\bibitem{ref_article16}
Vettoruzzo A, Bouguelia M R, Vanschoren J, et al. Advances and Challenges in Meta-Learning: A Technical Review[J]. arXiv preprint arXiv:2307.04722, 2023.

\bibitem{ref_article17}
Sun Y, Wang S, Feng S, et al. Ernie 3.0: Large-scale knowledge enhanced pre-training for language understanding and generation[J]. arXiv preprint arXiv:2107.02137, 2021.

\bibitem{ref_article18}
Deng J, Lin Y. The benefits and challenges of ChatGPT: An overview[J]. Frontiers in Computing and Intelligent Systems, 2022, 2(2): 81-83.

\bibitem{ref_article19}
Addington S. ChatGPT: Cyber Security Threats and Countermeasures[J]. Available at SSRN 4425678, 2023.

\bibitem{ref_article20}
Qammar A, Wang H, Ding J, et al. Chatbots to ChatGPT in a Cybersecurity Space: Evolution, Vulnerabilities, Attacks, Challenges, and Future Recommendations[J]. arXiv preprint arXiv:2306.09255, 2023.

\bibitem{ref_article21}
Charan P V, Chunduri H, Anand P M, et al. From Text to MITRE Techniques: Exploring the Malicious Use of Large Language Models for Generating Cyber Attack Payloads[J]. arXiv preprint arXiv:2305.15336, 2023.

\bibitem{ref_article22}
Shi J, Liu Y, Zhou P, et al. BadGPT: Exploring Security Vulnerabilities of ChatGPT via Backdoor Attacks to InstructGPT[J]. arXiv preprint arXiv:2304.12298, 2023.

\bibitem{ref_article23}
Gupta M, Akiri C K, Aryal K, et al. From ChatGPT to ThreatGPT: Impact of Generative AI in Cybersecurity and Privacy[J]. IEEE Access, 2023.

\bibitem{ref_article24}
Roy S S, Naragam K V, Nilizadeh S. Generating Phishing Attacks using ChatGPT[J]. arXiv preprint arXiv:2305.05133, 2023.

\bibitem{ref_article25}
Grbic D V, Dujlovic I. Social engineering with ChatGPT[C]//2023 22nd International Symposium INFOTEH-JAHORINA (INFOTEH). IEEE, 2023: 1-5.

\bibitem{ref_article26}
Dash, Bibhu, and Pawankumar Sharma. “Are ChatGPT and Deepfake Algorithms Endangering the Cybersecurity Industry? A Review.”  International Journal of Engineering and Applied Sciences, 2023, 10(1).

\bibitem{ref_article27}
Prasad S G, Sharmila V C, Badrinarayanan M K. Role of Artificial Intelligence based Chat Generative Pre-trained Transformer (ChatGPT) in Cyber Security[C]//2023 2nd International Conference on Applied Artificial Intelligence and Computing (ICAAIC). IEEE, 2023: 107-114.

\bibitem{ref_article28}
Rasheed H, Hadi A, Khader M. Threat hunting using grr rapid response[C]//2017 International Conference on New Trends in Computing Sciences (ICTCS). IEEE, 2017: 155-160.

\bibitem{ref_article29}
Sharma P, Dash B. Impact of big data analytics and ChatGPT on cybersecurity[C]//2023 4th International Conference on Computing and Communication Systems (I3CS). IEEE, 2023: 1-6.

\bibitem{ref_article30}
Dash, Bibhu, Pawankumar Sharma, and Azad Ali. "Federated Learning for Privacy-Preserving: A Review of PII Data Analysis in Fintech." International Journal of Software Engineering \& Applications (IJSEA) 13, no. 4, 2022.






\end{thebibliography}
\end{document}